\documentclass[conference]{IEEEtran}
\IEEEoverridecommandlockouts
\usepackage{cite}
\usepackage{amsmath,amssymb,amsfonts}
\usepackage{algorithmic}
\usepackage{graphicx}
\usepackage{textcomp}
\usepackage{xcolor}
\usepackage{units}
\usepackage{balance}
\usepackage{epsfig}
\usepackage{epstopdf}

\usepackage[ subrefformat=parens,labelformat=parens, caption=false, font=footnotesize]{subfig}

\graphicspath{{Figures/}}
\def\BibTeX{{\rm B\kern-.05em{\sc i\kern-.025em b}\kern-.08em
    T\kern-.1667em\lower.7ex\hbox{E}\kern-.125emX}}

\usepackage[colorinlistoftodos,bordercolor=orange,backgroundcolor=orange!20,linecolor=orange,textsize=scriptsize]{todonotes}
\makeatother

\makeatletter
\def\ps@IEEEtitlepagestyle{%
  \def\@oddfoot{\mycopyrightnotice}%
  \def\@oddhead{\hbox{}\@IEEEheaderstyle\leftmark\hfil\thepage}\relax
  \def\@evenhead{\@IEEEheaderstyle\thepage\hfil\leftmark\hbox{}}\relax
  \def\@evenfoot{}%
}

\def\mycopyrightnotice{%
  \begin{minipage}{\textwidth}
  \centering \scriptsize
    This work has been accepted by the 2024 IEEE 13th Sensor Array and Multichannel Signal Processing Workshop (SAM) for publication.  Copyright may be transferred without notice, after which this version may no longer be accessible.
  \end{minipage}
}
\makeatother

\begin{document}
%-------  Colors -------%
\newcommand{\red}[1]{{\color{red}{#1}}} 
\newcommand{\blue}[1]{{\color{blue}{#1}}} % for Simon 
\newcommand{\green}[1]{{\color{green}{#1}}} 
\newcommand{\yellow}[1]{{\color{yellow}{#1}}} 
\newcommand{\orange}[1]{{\color{orange}{#1}}} 

%-------  Mathematical functions -------%
\newcommand{\nth}[1]{{#1}{\text{th}}}
\newcommand{\mbf}[1]{\mathbf{#1}}

\newcommand{\Hpow}{{\sf H}}
\newcommand{\Tpow}{{\sf T}}
\newcommand{\Invpow}{{\sf -1}}
\newcommand{\Strpow}{{\sf *}}

\newcommand{\PseuInvpow}{\mathrm{\dagger}}
\newcommand{\Npow}{\mathrm{n}}
\newcommand{\abs}[1]{\left|{#1}\right|}
\newcommand{\norm}[1]{\left\|{#1}\right\|}
\newcommand{\vect}[1]{\mathrm{vec}\left(#1\right)}
\newcommand{\supp}[1]{\mathrm{supp}\left(#1\right)}
\newcommand{\trc}[1]{\mathrm{tr}\left(#1\right)}
\newcommand{\diagopr}[1]{\mathrm{diag}\left(#1\right)}
\newcommand{\blkdiagopr}[1]{\mathrm{blkdiag}\left(#1\right)}

\newcommand{\atantwo}{\mathrm{arctan2}}
%-------  Abbreviations -------%

\newcommand{\txidx}{\mathrm{T}}
\newcommand{\rxidx}{\mathrm{R}}
\newcommand{\strmidx}{\mathrm{S}}
\newcommand{\froidx}{\mathrm{F}}
\newcommand{\sysidx}{\mathrm{sys}}
\newcommand{\RFidx}{\mathrm{RF}}
\newcommand{\BBidx}{\mathrm{BB}}
\newcommand{\quantidx}{\mathrm{quant}}
\newcommand{\effidx}{\mathrm{eff}}
\newcommand{\totsupsc}{\mathrm{tot}}
\newcommand{\firstelement}{\mathrm{st}}

\newcommand{\pilotidx}{\mathrm{p}}
\newcommand{\Beamsupsc}{\mathrm{beam}}
\newcommand{\CP}{\mathrm{CP}}
\newcommand{\cent}{\mathrm{cen}}

\newcommand{\coh}{\mathrm{coh}}
\newcommand{\rms}{\mathrm{rms}}
\newcommand{\sym}{\mathrm{sym}}
\newcommand{\maxx}{\mathrm{dmax}}

\newcommand{\train}{\mathrm{tr}}
\newcommand{\ovs}{\mathrm{ovs}}
\newcommand{\DicRed}{\mathrm{RD}}
\newcommand{\LessCol}{\mathrm{lc}}

\newcommand{\clu}{\mathrm{clu}}
\newcommand{\ray}{\mathrm{ray}}
\newcommand{\LoS}{\mathrm{los}}
\newcommand{\NLoS}{\mathrm{nlos}}
\newcommand{\subb}{\mathrm{sub}}
\newcommand{\GMM}{\mathrm{GMM}}
\newcommand{\svv}{\mathrm{sv}}

\newcommand{\samp}{\mathrm{s}}
\newcommand{\ula}{\mathrm{ULA}}
\newcommand{\upa}{\mathrm{UPA}}
\newcommand{\bmsp}{\mathrm{bmsp}}

\newcommand{\NFsupsc}{\mathrm{NF}}
\newcommand{\FFsupsc}{\mathrm{FF}}
\newcommand{\SWMsupsc}{\mathrm{SWM}}
\newcommand{\HSPMsupsc}{\mathrm{HSPWM}}
\newcommand{\PWMsupsc}{\mathrm{PWM}}
\newcommand{\hmm}{\mathrm{HMM}}
\newcommand{\opt}{\mathrm{opt}}

\newcommand{\umarridx}{\mathrm{UMA}}
\newcommand{\saidx}{\mathrm{SA}}
\newcommand{\aeidx}{\mathrm{AE}}

\newcommand{\rotsupsc}{\mathrm{rot}}
\newcommand{\trialsupsc}{\mathrm{trl}}
\newcommand{\maxsupsc}{\mathrm{max}}
\newcommand{\Polsupsc}{\mathrm{pol}}
\newcommand{\DFTsupsc}{\mathrm{dft}}

\newcommand{\allidx}{\mathrm{all}}
\newcommand{\distsupsc}{\mathrm{dist}}
\newcommand{\offsupsc}{\mathrm{offline}}
\newcommand{\onsupsc}{\mathrm{online}}
\newcommand{\estidx}{\mathrm{est}}
%------- ... -------%
\setlength\unitlength{1mm}

\newcommand{\insertfig}[3]{
\begin{figure}[htbp]\begin{center}\begin{picture}(120,90)
\put(0,-5){\includegraphics[width=12cm,height=9cm,clip=]{#1.eps}}\end{picture}\end{center}
\caption{#2}\label{#3}\end{figure}}

\newcommand{\insertxfig}[4]{
\begin{figure}[htbp]
\begin{center}
\leavevmode \centerline{\resizebox{#4\textwidth}{!}{\input
#1.pstex_t}}
%\vspace*{-0.2in}
\caption{#2} \label{#3}
\end{center}
\end{figure}}

\long\def\comment#1{}

% bb font symbols

% \DeclareMathOperator*{\argmax}{arg\,max}
% \DeclareMathOperator*{\argmin}{arg\,min}

\newfont{\bbb}{msbm10 scaled 700}
\newcommand{\CCC}{\mbox{\bbb C}}

\newfont{\bb}{msbm10 scaled 1100}
\newcommand{\CC}{\mbox{\bb C}}
\newcommand{\PP}{\mbox{\bb P}}
\newcommand{\RR}{\mbox{\bb R}}
\newcommand{\QQ}{\mbox{\bb Q}}
\newcommand{\ZZ}{\mbox{\bb Z}}
\newcommand{\FF}{\mbox{\bb F}}
\newcommand{\GG}{\mbox{\bb G}}
\newcommand{\EE}{\mbox{\bb E}}
\newcommand{\NN}{\mbox{\bb N}}
\newcommand{\KK}{\mbox{\bb K}}

% Vectors

\newcommand{\av}{{\bf a}}
\newcommand{\bv}{{\bf b}}
\newcommand{\cv}{{\bf c}}
\newcommand{\dv}{{\bf d}}
\newcommand{\ev}{{\bf e}}
\newcommand{\fv}{{\bf f}}
\newcommand{\gv}{{\bf g}}
\newcommand{\hv}{{\bf h}}
\newcommand{\iv}{{\bf i}}
\newcommand{\jv}{{\bf j}}
\newcommand{\kv}{{\bf k}}
\newcommand{\lv}{{\bf l}}
\newcommand{\mv}{{\bf m}}
\newcommand{\nv}{{\bf n}}
\newcommand{\ov}{{\bf o}}
\newcommand{\pv}{{\bf p}}
\newcommand{\qv}{{\bf q}}
\newcommand{\rv}{{\bf r}}
\newcommand{\sv}{{\bf s}}
\newcommand{\tv}{{\bf t}}
\newcommand{\uv}{{\bf u}}
\newcommand{\wv}{{\bf w}}
\newcommand{\xv}{{\bf x}}
\newcommand{\yv}{{\bf y}}
\newcommand{\zv}{{\bf z}}
\newcommand{\zerov}{{\bf 0}}
\newcommand{\onev}{{\bf 1}}

\def\u{{\bf u}}

% Matrices

\newcommand{\Am}{{\bf A}}
\newcommand{\Bm}{{\bf B}}
\newcommand{\Cm}{{\bf C}}
\newcommand{\Dm}{{\bf D}}
\newcommand{\Em}{{\bf E}}
\newcommand{\Fm}{{\bf F}}
\newcommand{\Gm}{{\bf G}}
\newcommand{\Hm}{{\bf H}}
\newcommand{\Id}{{\bf I}}
\newcommand{\Jm}{{\bf J}}
\newcommand{\Km}{{\bf K}}
\newcommand{\Lm}{{\bf L}}
\newcommand{\Mm}{{\bf M}}
\newcommand{\Nm}{{\bf N}}
\newcommand{\Om}{{\bf O}}
\newcommand{\Pm}{{\bf P}}
\newcommand{\Qm}{{\bf Q}}
\newcommand{\Rm}{{\bf R}}
\newcommand{\Sm}{{\bf S}}
\newcommand{\Tm}{{\bf T}}
\newcommand{\Um}{{\bf U}}
\newcommand{\Wm}{{\bf W}}
\newcommand{\Vm}{{\bf V}}
\newcommand{\Xm}{{\bf X}}
\newcommand{\Ym}{{\bf Y}}
\newcommand{\Zm}{{\bf Z}}
\newcommand{\Onem}{{\bf 1}}
\newcommand{\Zerom}{{\bf 0}}

% Calligraphic

\newcommand{\Ac}{{\cal A}}
\newcommand{\Bc}{{\cal B}}
\newcommand{\Cc}{{\cal C}}
\newcommand{\Dc}{{\cal D}}
\newcommand{\Ec}{{\cal E}}
\newcommand{\Fc}{{\cal F}}
\newcommand{\Gc}{{\cal G}}
\newcommand{\Hc}{{\cal H}}
\newcommand{\Ic}{{\cal I}}
\newcommand{\Jc}{{\cal J}}
\newcommand{\Kc}{{\cal K}}
\newcommand{\Lc}{{\cal L}}
\newcommand{\Mc}{{\cal M}}
\newcommand{\Nc}{{\cal N}}
\newcommand{\Oc}{{\cal O}}
\newcommand{\Pc}{{\cal P}}
\newcommand{\Qc}{{\cal Q}}
\newcommand{\Rc}{{\cal R}}
\newcommand{\Sc}{{\cal S}}
\newcommand{\Tc}{{\cal T}}
\newcommand{\Uc}{{\cal U}}
\newcommand{\Wc}{{\cal W}}
\newcommand{\Vc}{{\cal V}}
\newcommand{\Xc}{{\cal X}}
\newcommand{\Yc}{{\cal Y}}
\newcommand{\Zc}{{\cal Z}}

% Bold greek letters

\newcommand{\alphav}{\hbox{\boldmath$\alpha$}}
\newcommand{\betav}{\hbox{\boldmath$\beta$}}
\newcommand{\gammav}{\hbox{\boldmath$\gamma$}}
\newcommand{\deltav}{\hbox{\boldmath$\delta$}}
\newcommand{\etav}{\hbox{\boldmath$\eta$}}
\newcommand{\lambdav}{\hbox{\boldmath$\lambda$}}
\newcommand{\epsilonv}{\hbox{\boldmath$\epsilon$}}
\newcommand{\nuv}{\hbox{\boldmath$\nu$}}
\newcommand{\muv}{\hbox{\boldmath$\mu$}}
\newcommand{\zetav}{\hbox{\boldmath$\zeta$}}
\newcommand{\phiv}{\hbox{\boldmath$\phi$}}
\newcommand{\psiv}{\hbox{\boldmath$\psi$}}
\newcommand{\thetav}{\hbox{\boldmath$\theta$}}
\newcommand{\tauv}{\hbox{\boldmath$\tau$}}
\newcommand{\omegav}{\hbox{\boldmath$\omega$}}
\newcommand{\xiv}{\hbox{\boldmath$\xi$}}
\newcommand{\sigmav}{\hbox{\boldmath$\sigma$}}
\newcommand{\piv}{\hbox{\boldmath$\pi$}}
\newcommand{\rhov}{\hbox{\boldmath$\rho$}}
\newcommand{\vtv}{\hbox{\boldmath$\vartheta$}}

\newcommand{\Gammam}{\hbox{\boldmath$\Gamma$}}
\newcommand{\Lambdam}{\hbox{\boldmath$\Lambda$}}
\newcommand{\Deltam}{\hbox{\boldmath$\Delta$}}
\newcommand{\Sigmam}{\hbox{\boldmath$\Sigma$}}
\newcommand{\Phim}{\hbox{\boldmath$\Phi$}}
\newcommand{\Pim}{\hbox{\boldmath$\Pi$}}
\newcommand{\Psim}{\hbox{\boldmath$\Psi$}}
\newcommand{\psim}{\hbox{\boldmath$\psi$}}
\newcommand{\chim}{\hbox{\boldmath$\chi$}}
\newcommand{\omegam}{\hbox{\boldmath$\omega$}}
\newcommand{\vphim}{\hbox{\boldmath$\varphi$}}
\newcommand{\Thetam}{\hbox{\boldmath$\Theta$}}
\newcommand{\Omegam}{\hbox{\boldmath$\Omega$}}
\newcommand{\Xim}{\hbox{\boldmath$\Xi$}}

% mixed symbols

\newcommand{\sinc}{{\hbox{sinc}}}
\newcommand{\diag}{{\hbox{diag}}}
\renewcommand{\det}{{\hbox{det}}}
\newcommand{\trace}{{\hbox{tr}}}
\newcommand{\sign}{{\hbox{sign}}}
\renewcommand{\arg}{{\hbox{arg}}}
\newcommand{\var}{{\hbox{var}}}
\newcommand{\cov}{{\hbox{cov}}}
\newcommand{\SINR}{{\sf sinr}}
\newcommand{\SNR}{{\sf snr}}
\newcommand{\Ei}{{\rm E}_{\rm i}}
\newcommand{\eqdef}{\stackrel{\Delta}{=}}
\newcommand{\defines}{{\,\,\stackrel{\scriptscriptstyle \bigtriangleup}{=}\,\,}}
\newcommand{\<}{\left\langle}
\renewcommand{\>}{\right\rangle}
\newcommand{\herm}{{\sf H}}
\newcommand{\trasp}{{\sf T}}
\renewcommand{\vec}{{\rm vec}}
\newcommand{\calL}{\mbox{${\mathcal L}$}}
\newcommand{\calO}{\mbox{${\mathcal O}$}}

\newcommand{\Afd}{\mbox{$\boldsymbol{\mathcal{A}}$}}
\newcommand{\Bfd}{\mbox{$\boldsymbol{\mathcal{B}}$}}
\newcommand{\Cfd}{\mbox{$\boldsymbol{\mathcal{C}}$}}
\newcommand{\Dfd}{\mbox{$\boldsymbol{\mathcal{D}}$}}
\newcommand{\Efd}{\mbox{$\boldsymbol{\mathcal{E}}$}}
\newcommand{\Ffd}{\mbox{$\boldsymbol{\mathcal{F}}$}}
\newcommand{\Gfd}{\mbox{$\boldsymbol{\mathcal{G}}$}}
\newcommand{\Hfd}{\mbox{$\boldsymbol{\mathcal{H}}$}}
\newcommand{\Ifd}{\mbox{$\boldsymbol{\mathcal{I}}$}}
\newcommand{\Jfd}{\mbox{$\boldsymbol{\mathcal{J}}$}}
\newcommand{\Kfd}{\mbox{$\boldsymbol{\mathcal{K}}$}}
\newcommand{\Lfd}{\mbox{$\boldsymbol{\mathcal{L}}$}}
\newcommand{\Mfd}{\mbox{$\boldsymbol{\mathcal{M}}$}}
\newcommand{\Nfd}{\mbox{$\boldsymbol{\mathcal{N}}$}}
\newcommand{\Ofd}{\mbox{$\boldsymbol{\mathcal{O}}$}}
\newcommand{\Pfd}{\mbox{$\boldsymbol{\mathcal{P}}$}}
\newcommand{\Qfd}{\mbox{$\boldsymbol{\mathcal{Q}}$}}
\newcommand{\Rfd}{\mbox{$\boldsymbol{\mathcal{R}}$}}
\newcommand{\Sfd}{\mbox{$\boldsymbol{\mathcal{S}}$}}
\newcommand{\Tfd}{\mbox{$\boldsymbol{\mathcal{T}}$}}
\newcommand{\Ufd}{\mbox{$\boldsymbol{\mathcal{U}}$}}
\newcommand{\Vfd}{\mbox{$\boldsymbol{\mathcal{V}}$}}
\newcommand{\Wfd}{\mbox{$\boldsymbol{\mathcal{W}}$}}
\newcommand{\Xfd}{\mbox{$\boldsymbol{\mathcal{X}}$}}
\newcommand{\Yfd}{\mbox{$\boldsymbol{\mathcal{Y}}$}}
\newcommand{\Zfd}{\mbox{$\boldsymbol{\mathcal{Z}}$}}

\title{Near or far: On determining the appropriate channel estimation strategy in cross-field communication\\
\thanks{This work was supported by the KAUST Office of Sponsored Research.}
}

\author{\IEEEauthorblockN{Simon~Tarboush}
\IEEEauthorblockA{\textit{Independent Researcher,} \\
Dubai, UAE.\\
simon.w.tarboush@gmail.com}
\and
\IEEEauthorblockN{Anum Ali}
\IEEEauthorblockA{\textit{Standards and Mobility Innovation Lab,} \\
\textit{Samsung Research America,}\\
Texas, USA. \\
anum.ali@samsung.com}
\and
\IEEEauthorblockN{Tareq~Y.~Al-Naffouri}
\IEEEauthorblockA{\textit{Department of CEMSE, Information Science Lab,} \\ % Department of CEMSE - 
\textit{King Abdullah University of Science and Technology,}\\
Thuwal, KSA. \\
tareq.alnaffouri@kaust.edu.sa}
}

\maketitle

\begin{abstract}
The use of ultra-massive multiple-input multiple-output and high-frequency large bandwidth systems is likely in the next-generation wireless communication systems. In such systems, the user moves between near- and far-field regions, and consequently, the channel estimation will need to be carried out in the cross-field scenario. Channel estimation strategies have been proposed for both near- and far-fields, but in the cross-field problem, the first step is to determine whether the near- or far-field is applicable so that an appropriate channel estimation strategy can be employed. In this work, we propose using a hidden Markov model over an ensemble of region estimates to enhance the accuracy of selecting the actual region. The region indicators are calculated using the pair-wise power differences between received signals across the subarrays within an array-of-subarrays architecture. Numerical results show that the proposed method achieves a high success rate in determining the appropriate channel estimation strategy.
\end{abstract}


\section{Introduction}
\label{sec:intro}
By combining ultra-massive multiple-input multiple-output (UM-MIMO) with large bandwidths, future wireless communication systems promise substantial improvement in spectral efficiency and seamless communication with multiple users in dense environments~\cite{lu2021communicating,ramezani2023exploiting,wang2024tutorial,bjornson2024towards}.

The distinctive characteristics of UM-MIMO and high-frequency channels offer opportunities but also pose challenges in channel modeling and estimation~\cite{sarieddeen2021overview}. These challenges arise from fundamental changes in electromagnetic characteristics, e.g., when communicating in the near-field~\cite{cui2022near}. Furthermore, the system design becomes intricate due to large antenna arrays and bandwidth. Consequently, the hybrid beamforming sub-connected array-of-subarrays (AoSA) architecture and its variants~\cite{song2019fully,tarboush2021teramimo,han2021hybrid} have emerged as promising solutions. Sub-connected AoSA architecture respects hardware and power constraints while simplifying channel estimation~\cite{tarboush2023compressive}.

The accurate channel model in any region, including the near-field, is the spherical wave model (SWM)~\cite{cui2022near}, while the planar wave model (PWM) is accurate enough for the far-field. An alternative model tailored for the AoSA architecture is the hybrid spherical-planar wave model (HSPWM)~\cite{tarboush2023compressive} where the propagation is characterized by PWM within a subarray (SA), while the SWM is used across SAs. Generally, the HSPWM model provides a reasonable compromise between modeling complexity and accuracy. For small distances, however, channel estimation based on HSPWM is inferior in performance to the SWM system~\cite{tarboush2023cross}. Thus, the problem lies in determining whether HSPWM or SWM should be used in the online operation, i.e., whether near- or far-field assumption should be made while estimating the channel of one SA within the AoSA system. This channel estimation problem across the near- and far-field is called ``Cross-field channel estimation"~\cite{tarboush2023cross,han2023cross,chen2023can}. Channel estimation methods have been proposed separately for HSPWM~\cite{tarboush2023compressive,chen2021hybrid,sun2023subarray,zhu2023sub} and SWM~\cite{cui2022channel,lu2023near}, but using the same estimation method regardless of the distance is sub-optimal~\cite{tarboush2023cross}. Specifically, using an appropriate estimation method can lead to a better trade-off in computational complexity and estimation accuracy~\cite{tarboush2023cross}.

Essentially, a practical method for online determination of the appropriate propagation model and, consequently, the estimation strategy is required. The challenge lies in determining a suitable estimation method based solely on the received signals before initiating the estimation procedure itself, and in the absence of any information on the relative location between the transmitter and the receiver. In this work, we elaborate on and do a thorough evaluation of a ``channel estimation method" selection metric that was proposed for this purpose in~\cite{tarboush2023cross}. The proposed metric uses the variation of received signals across the SAs to determine a suitable channel estimation method. Note that the received signals are the sole information available at this stage. Further, we introduce and evaluate a hidden Markov model (HMM) for selection with a series of observations. We showed that the HMM-based approach has a substantially better decision accuracy compared to the baseline~\cite{tarboush2023cross}.

\textbf{Notation:} Non-bold lower and upper case letters $a, A$ denote scalars, bold lower case letters $\mbf{a}$ denote vector, and bold upper case letters $\mbf{A}$ denote matrices. $\mbf{I}_N$ is the identity matrix of size size $N\times N$. $\abs{\cdot}$ denotes the absolute value of $a$ and absolute value of each entry of $\mbf{a}$. $\norm{\mbf{A}}_F$ is the Frobenius norm and $\vect{\mbf{A}}$ is the vectorized version of $\mbf{A}$ obtained by stacking the column of $\mbf{A}$. $j=\sqrt{-1}$ denotes the imaginary unit. The superscripts ${(\cdot)}^\Tpow$ and ${(\cdot)}^\Hpow$ stand for the transpose and conjugate transpose, respectively, and the subscripts $(\cdot)_{\txidx}$ and $(\cdot)_{\rxidx}$ denote the transmitter (Tx) and receiver (Rx), respectively.
\section{System and Channel Model}
\label{sec:System_Channel_CrossField}
\subsection{AoSA Architecture}
\label{sec:System_Model}
\begin{figure}[htb]
  \centering
  \includegraphics[width = 0.85\linewidth]{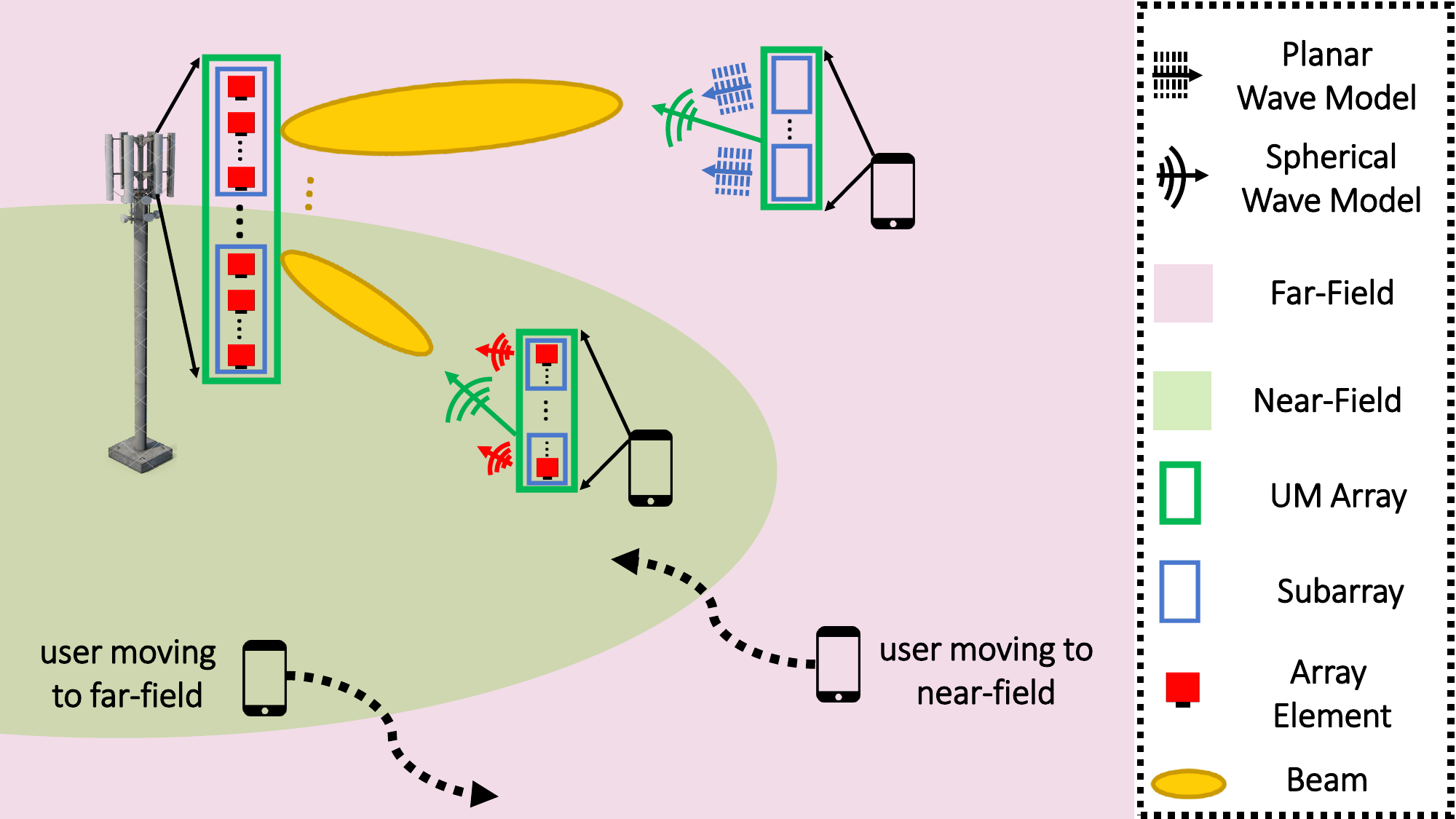}
  \caption{Illustration of the cross-field channel estimation problem, depicting various channel models suitable for different distances. The transmitter and receiver employ an UM-MIMO AoSA architecture in a sub-THz/THz-band communication system.}
  \label{fig:aosa_tx_rx}
  \vspace{-2mm}
\end{figure}

Consider a downlink multi-carrier sub-THz/THz band system~\cite{tarboush2022single}, as illustrated in Fig.~\ref{fig:aosa_tx_rx}, employing UM arrays at both the Tx and Rx. The system dimension is $N_\rxidx\times N_\txidx$, and it operates over an ultra-wide bandwidth $B_\sysidx$ spanning $K$ subcarriers. The frequency of the $\nth{k}$ subcarrier is $f_k=f_c+\frac{B_\sysidx}{K}(k-\frac{K-1}{2}), k =\{1,\cdots,K\}$, where $f_c$ is the center frequency. The system deploys an AoSA architecture, where each SA contains multiple array elements (AEs) and is connected exclusively to one radio frequency (RF)-chain. The Tx is equipped with $Q_\txidx$ SAs, and each Tx SA comprises a uniform linear array (ULA) with $\bar{Q}_\txidx$ closely spaced AEs separated by $\delta_\txidx$. The centers of Tx SAs are spaced apart by $\Delta_\txidx$. A similar architecture is used at the Rx~\cite{tarboush2021teramimo}.

The Rx signal at the $\nth{k}$ subcarrier can be expressed as
\begin{equation}
    \mbf{y}[k]=\mbf{W}^\Hpow_{\BBidx}[k]\mbf{W}^\Hpow_{\RFidx}\left(\mbf{H}[k]\mbf{F}_{\RFidx}\mbf{F}_{\BBidx}[k]\mbf{s}[k]+\mbf{n}[k]\right),
    \label{eq:rx_subc_sig}
\end{equation}
where the Tx vector $\mbf{s}[k]\in\CC^{N_{\strmidx}\times1}$ represents pilot or data streams with $N_{\strmidx}\leq \min\{Q_\txidx,Q_\rxidx\}$, $\mbf{F}_{\BBidx}[k]/\mbf{W}_{\BBidx}[k]$ denotes the baseband digital precoder/combiner, the analog RF beamformers/combiners $\mbf{F}_{\RFidx}[k]/\mbf{W}_{\RFidx}[k]$ have constant magnitude and adopt finite-resolution wideband phase-shifters following large bandwidth hardware constraints~\cite{tarboush2023cross}, each matrix has block diagonal structure, and $\mbf{n}[k]\sim\mathcal{CN}\left(\mbf{0},\sigma_n^2\mbf{I}_{N_\rxidx}\right)$ is the additive white Gaussian noise vector, where $\sigma_n^2$ is the noise power. The UM-MIMO channel matrix $\mbf{H}[k]$ consists of all frequency-domain sub-channels $\mbf{H}_{q_\rxidx,q_\txidx}[k]\in\CC^{\bar{Q}_\rxidx\times\bar{Q}_\txidx}$ defined between the $\nth{q_\txidx}$ Tx SA and the $\nth{q_\rxidx}$ Rx SA~\cite{tarboush2021teramimo}. The hybrid precoders (combiners) satisfy the total power constraint $\sum_{k=1}^{K}\left(\norm{\mbf{F}_\RFidx\mbf{F}_\BBidx[k]}_{\froidx}^2\right)=KN_{\strmidx}$. The channel coherence time is $T_{\coh}\!=\!\sqrt{\frac{9}{16\pi f_{\maxx}}\times\frac{1}{f_{\maxx}}}$~\cite{tarboush2021teramimo}, where $f_{\maxx}\!=\!\frac{v}{c_0}f_c$ is the maximum Doppler shift, $v$ is the velocity of the Tx or Rx, in $\unit{(m/sec)}$, and $c_0$ is the speed of light.
\subsection{Channel Models}
\label{sec:Channel_Model}
\subsubsection{SWM}
\label{sec:swm_Channel_Model}
The ground truth and the most accurate model for channel modeling at any communication distance is the SWM. The SWM individually models the channel between all Tx-Rx AE pairs, capturing both the magnitude and phase variations of the received signal across the AEs. It is particularly suitable for near-field. Consequently, the near-field MIMO channel between $\nth{q_\txidx}$ Tx SA and $\nth{q_\rxidx}$ Rx SA at subcarrier $k$ is~\cite{tarboush2023cross}
\begin{equation}  
\label{eq:H_SWM_withapproximation}
    \begin{split}
    \vspace{-2mm}
    \mbf{H}_{q_\rxidx,q_\txidx}^\SWMsupsc[k]=&\sqrt{\frac{\bar{Q}_\rxidx\bar{Q}_\txidx}{L}}\sum_{\ell=1}^{L}\!\alpha^\ell(f_k,d^\ell_{(n_\rxidx,n_\txidx)})\times\\
    &\mbf{b}_\rxidx(\phi^\ell,d^\ell_\rxidx)\mbf{b}^\Tpow_\txidx(\theta^\ell,d^\ell_\txidx),
    \end{split}
\end{equation}
where $L$ represents the number of paths, $\alpha^\ell$ denotes the complex channel gain of the $\nth{\ell}$ path, $d^\ell_{(n_\rxidx,n_\txidx)}$ is the length of the $\nth{\ell}$ propagation path between the $\nth{n_\txidx}$ Tx AE and $\nth{n_\rxidx}$ Rx AE, $\mbf{b}(\theta,d)$  is the near-field array response vector (ARV) - a function of both angle and distance~\cite{cui2022channel,tarboush2023cross}, and $\theta/\phi\in[0,\pi]$ denotes the angle-of-departure (AoD)/angle-of-arrival (AoA). In summary, as indicated by~\eqref{eq:H_SWM_withapproximation}, the distances, AoAs, and AoDs vary across the AEs.
\subsubsection{HSPWM}
\label{sec:HSPWM_Channel_Model}
In an UM-MIMO AoSA, the HSPWM captures only the phase variation across the adjacent AEs - based on PWM - since the SA usually contains significantly fewer AEs than the entire UM array, whereas both the phase and magnitude variations across spatially separated SAs are captured by SWM. HSPWM achieves a trade-off between modeling accuracy and complexity compared to SWM and PWM. Consequently, all MIMO channels between Tx SA $q_\txidx$ and Rx SA $q_\rxidx$ at subcarrier $k$ are defined by
\begin{equation}
\label{eq:ch_HSPWM_frequencydomain_vectnot}
    \begin{split}
    \mbf{H}_{q_\rxidx,q_\txidx}^\HSPMsupsc[k]\!=&\sqrt{\frac{\bar{Q}_\rxidx\bar{Q}_\txidx}{L}}\!\sum_{\ell=1}^{L}\alpha^\ell(f_k,d_{q_\rxidx,q_\txidx}^\ell)e^{-j\frac{2\pi}{\lambda_k}d^\ell_{q_\rxidx,q_\txidx}}\!\times\!\\&\mbf{a}_\rxidx\left(\phi^\ell_{q_\txidx,q_\rxidx}\right)\mbf{a}_\txidx^\Tpow\left(\theta^\ell_{q_\rxidx,q_\txidx}\right),
    \end{split}
\end{equation}
where $\lambda_k=c_0/f_k$ is the wavelength and $\mbf{a}(\theta)$ is the far-field ARV, a function of angle~\cite{tarboush2023cross,rodriguez2018frequency}. Note that - based on~\eqref{eq:ch_HSPWM_frequencydomain_vectnot} - distances, AoAs, and AoDs vary across the SAs. The HSPWM-based AoSA channel~\eqref{eq:ch_HSPWM_frequencydomain_vectnot} is different from the SWM~\eqref{eq:H_SWM_withapproximation}, as the near-field ARV $\mbf{b}(\theta,d)$ is replaced by the far-field ARV $\mbf{a}(\theta)$. Consequently, each scenario necessitates an appropriate channel estimation strategy since the unknown parameters vary across channel models. 
\section{Proposed Model Selection Strategy}
\label{sec:appr_ch_sel}
The cross-field problem is conceptualized as a classification task, aiming to decide whether to employ a near- or far-field channel estimation approach. The proposed strategy involves two main phases: $(\mathrm{i})$ collecting multiple measurements of a selection metric, denoted as $\eta$, and $(\mathrm{ii})$ applying a HMM to an ensemble of decisions.

\textbf{Phase 1:} The initial step involves a beam training procedure using pre-computed random beam weights. During this phase, for a specific pair of Tx-Rx measurements $(m_\txidx, m_\rxidx)$, the beamforming weight $\mbf{z}_{m_\txidx}$ is applied to the reference Tx SA. Simultaneously, the same combining weight vector $\mbf{c}_{m_\rxidx}$ is applied to all Rx SAs. This approach allows to capture the received signal variations across SAs resulting from the channel. The received signal, after a total of $M_\rxidx\times M_\txidx$ training beams, is~\cite{tarboush2023cross}
\begin{equation}
    \mbf{Y}^{q_\rxidx,q_\txidx}[k]=\sqrt{P_\txidx}\mbf{C}^\Hpow\mbf{H}^\SWMsupsc_{q_\rxidx,q_\txidx}[k]\mbf{Z}+\mbf{C}^\Hpow\mbf{N}[k],
    \label{eq:allmeasure_trainingphase_rxsig_subck}
\end{equation}
where $P_\txidx$ is the total Tx power per transmission during the training phase, $\mbf{C}={\left[{\mbf{c}_1,\mbf{c}_2,\cdots,\mbf{c}_{M_\rxidx}}\right]}^\Tpow$ is the $\bar{Q}_\rxidx\times M_\rxidx$ combining matrix used at the $\nth{q_\rxidx}$ Rx SA, $\mbf{Z}={\left[{\mbf{z}_1,\mbf{z}_2,\cdots,\mbf{z}_{M_\txidx}}\right]}^\Tpow$ is the $\bar{Q}_\txidx\times M_\txidx$ beamforming matrix used at the $\nth{q_\txidx}$ Tx SA, and $\mbf{N}[k]$ is the noise matrix.

The main idea behind the selection metric $\eta$ is to capture the variation in the received power across all Rx SAs. Formally, $\eta$ is defined as
\begin{equation}
    \label{eq:eta_metric}
    \begin{split}
    \eta&=\max{\{[\eta_{1,2},\!\cdots\!,\eta_{r,c},\!\cdots\!,\eta_{Q_\rxidx-1,Q_\rxidx}]\}}\quad\mathrm{with}\\&
    \eta_{r,c} = (\boldsymbol{\chi}^{q_r,1}-\boldsymbol{\chi}^{q_c,1})^\Hpow(\boldsymbol{\chi}^{q_r,1}-\boldsymbol{\chi}^{q_c,1})
    \end{split}
\end{equation}
where $r=\{1,\cdots,Q_\rxidx-1\}$, $c=\{r+1,\cdots,Q_\rxidx\}$, and the normalized vector is $\boldsymbol{\chi}^{q_r,1}=\frac{\abs{\bar{\boldsymbol{\chi}}^{q_r,1}}}{\norm{\bar{\boldsymbol{\chi}}^{q_r,1}}}$ where $\bar{\boldsymbol{\chi}}^{q_r,1}=[\vect{\mbf{Y}^{q_\rxidx,1}[1]}^\Tpow,\cdots,\vect{\mbf{Y}^{q_\rxidx,1}[K]}^\Tpow]^\Tpow$ is obtained after collecting the vectorized version of the received signal in~\eqref{eq:allmeasure_trainingphase_rxsig_subck} for all subcarriers. It is important to emphasize that conducting beam training solely for the reference Tx SA is sufficient to determine which channel estimation method should be used. The selection metric is a function of training measurements and inherently takes into account the noise levels and the number of channel paths. Intuitively, we expect a lot of variation in received signals across SAs for near-field, and only a slight - if any - variation in far-field. Hence, $\eta$ exhibits a decreasing trend with distance.

The decisions about the appropriate region are derived using the following rule
\begin{equation}
\label{eq:decision_rule}
    \Omega=\left\{\begin{array}{ll}
    \mathrm{swm} & \mathrm{if} \quad\eta\geq\gamma_{\mathrm{N}\text{-}\mathrm{F}} \\
    \mathrm{hspwm}& \mathrm{else},
\end{array}\right.
\end{equation}
where the threshold $\gamma_{\mathrm{N}\text{-}\mathrm{F}}$ is used to distinguish between near and far fields. Note that the threshold $\gamma_{\mathrm{N}\text{-}\mathrm{F}}$ that represents the region boundary is derived by training across distances spanning near- and far-fields. The process involves calculating the metric $\eta_\offsupsc$ for each distance, with multiple orientations and trials $R_\offsupsc\times E_\offsupsc$, see~\cite{tarboush2023cross} for details.

\textbf{Phase 2:} As the user moves with a velocity $v$, the beam training procedure is repeated frequently. Therefore, we can collect multiple decisions, by applying \eqref{eq:decision_rule} to $\eta_\onsupsc$, in the observation vector $\mbf{O}=\{O_1, O_2, \cdots,O_T\}$ where $T$ is the total number of observations in the sequence. The decision about the region can be made by using $U$ most recent observations and performing a majority vote on the observations. Note that, the special case of $U=1$, i.e., using only the most recent observation was used in~\cite{tarboush2023cross}. Furthermore, to enhance the decision accuracy we propose to employ a HMM $\lambda_\hmm = (\mbf{A},\mbf{B},\boldsymbol{\pi})$ where $\mbf{A}$ is the transition probability matrix, $\mbf{B}$ is the observation probability matrix, and $\boldsymbol{\pi}$ is the initial state distribution~\cite{elliott2008hidden}. The possible hidden states for the model are $\mbf{S}=\{\mathrm{Near},\mathrm{Far}\}$ and each state can emit two possible observations $O_s \in\{\mathrm{swm},\mathrm{hspwm}\}$ with different probabilities following $\mbf{B}$. To obtain the model parameters $\lambda_\hmm$, an initial estimation is required which can be achieved following the same procedure as getting the offline threshold~\cite{tarboush2023cross}. In the online operation, the Rx collects the sequence $\mbf{O}_\onsupsc=\{O_{u-U+1},\cdots,O_u\}$ and the Viterbi algorithm is used to calculate the most likely path through this sequence\cite{lou1995implementing}.
\section{Simulation Results and Discussion}
\label{sec:sim_discc}
\begin{figure*}[htb]
  \centering
  \resizebox{0.77\textwidth}{!}{%
  \subfloat[Changing the number of Rx SAs and AEs with a fixed Rx aperture.]
  {\label{fig:res1a} \includegraphics[width=0.5\linewidth]{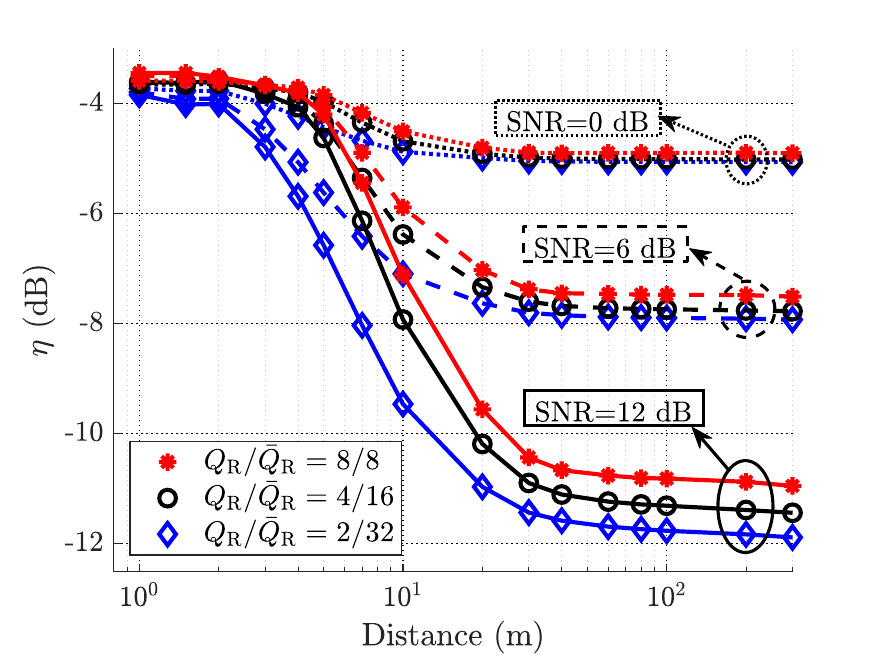}}
  \hfill
  \subfloat[Changing the number of Tx SAs and AEs with a fixed Tx aperture.]
  {\label{fig:res1b} \includegraphics[width=0.5\linewidth]{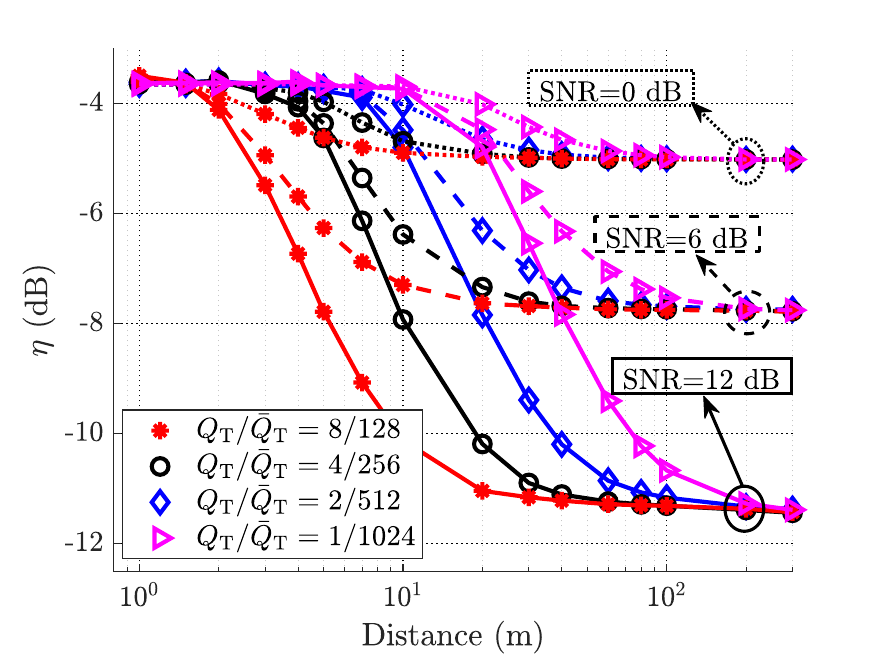}}%
  }
  \resizebox{0.77\textwidth}{!}{%
  \subfloat[Changing the Rx aperture.]
  {\label{fig:res1c} \includegraphics[width=0.5\linewidth]{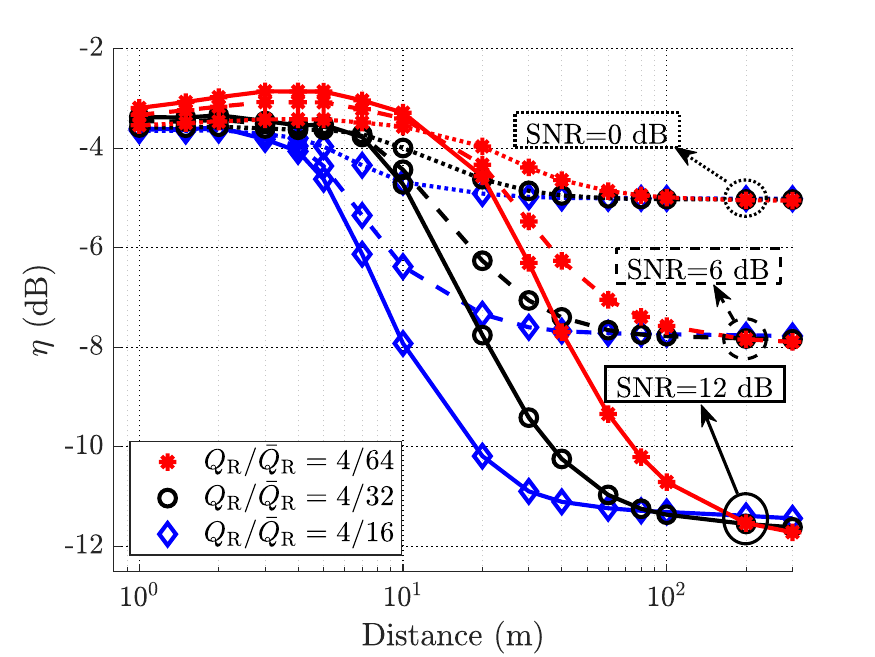}}
  \hfill
  \subfloat[Changing the Tx aperture.]
  {\label{fig:res1d} \includegraphics[width=0.5\linewidth]{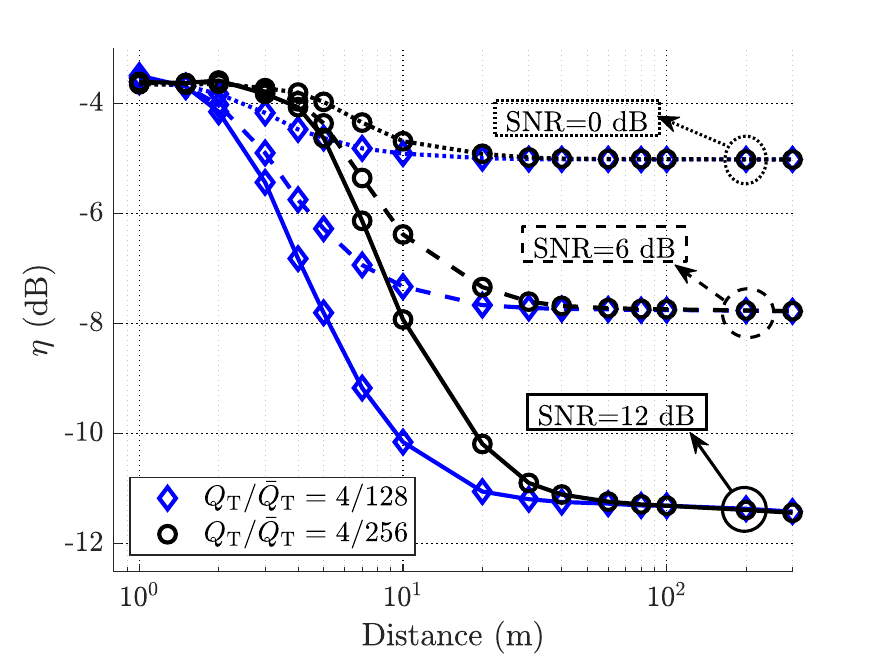}}%
  }
 \caption{Analysis of the proposed selection metric $\eta$ versus distance in the offline training stage for various $N_\txidx \times N_\rxidx$ Tx-Rx configurations and different SNR values: (a) $1024 \times 64$ with a Tx uses $Q_\txidx/\bar{Q}_\txidx = 4/256$ SAs/AEs, (b) $1024 \times 64$ with a Rx uses $Q_\rxidx/\bar{Q}_\rxidx = 4/16$ SAs/AEs, (c) changing $N_\rxidx$ by increasing the Rx AEs within each Rx SAs for a fixed Tx configuration with $N_\txidx=1024$ and $Q_\txidx/\bar{Q}_\txidx = 4/256$ SAs/AEs, and (d) changing $N_\txidx$ by increasing the Tx AEs within each Tx SAs for a fixed Rx configuration with $N_\rxidx=64$ and $Q_\rxidx/\bar{Q}_\rxidx = 4/16$ SAs/AEs.}
 \label{fig:prop_eta_ana}
  \vspace{-2mm}
 \centering
 \resizebox{0.76\textwidth}{!}{%
  \subfloat[Near-field region.]
  {\label{fig:resmv1} \includegraphics[width=0.5\linewidth]{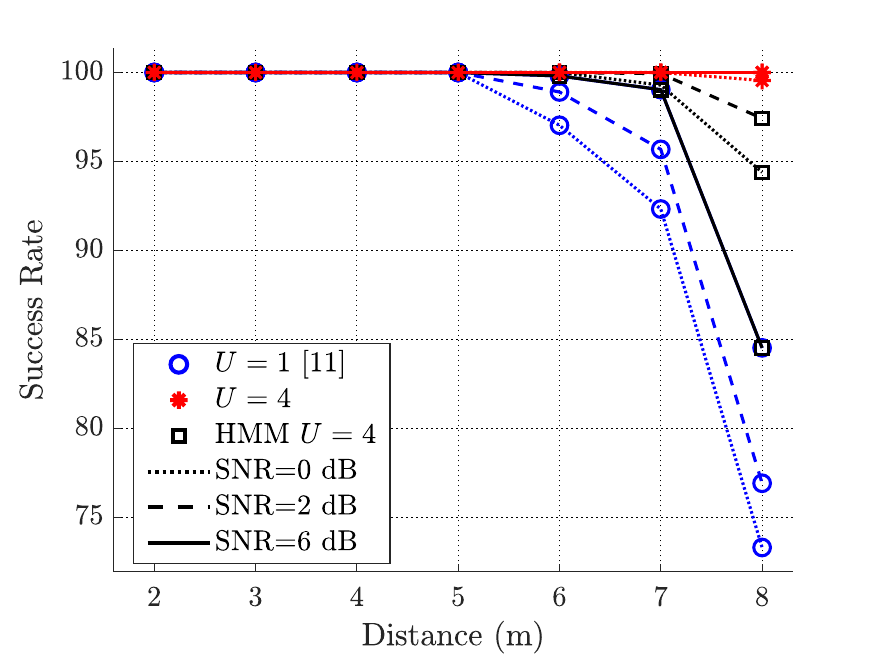}}
  \hfill
  \subfloat[Far-field region.]
  {\label{fig:resmv2} \includegraphics[width=0.5\linewidth]{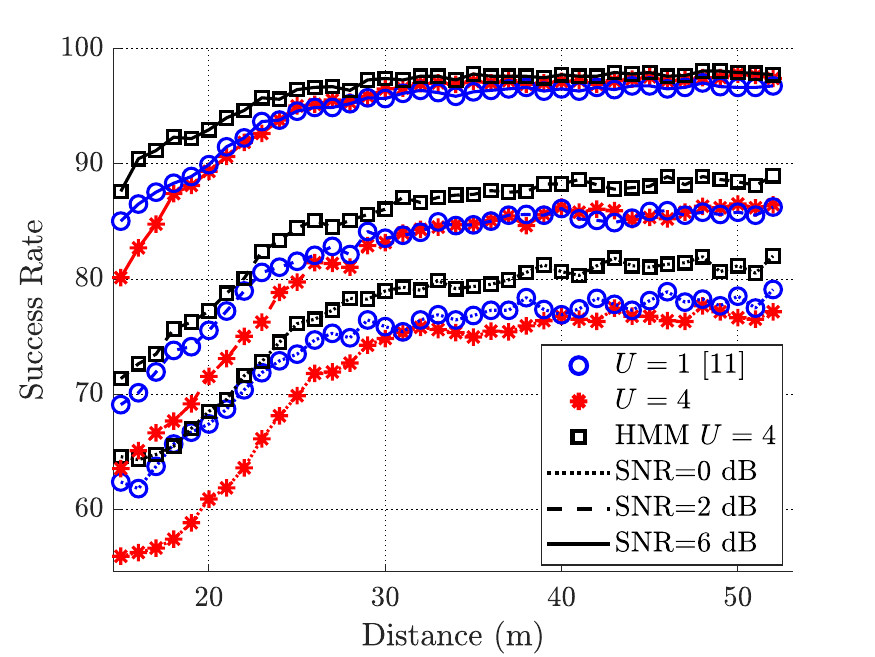}}%
  }
 \caption{Comparing the decision accuracy of the hidden Markov model $U=4$, majority vote $U=4$, and $U=1$~\cite{tarboush2023cross} across the near- and far-fields for different SNR values.}
 \label{fig:SV_vs_MV}
 \vspace{-5mm}
\end{figure*}

The default simulation settings are listed in Table~\ref{table:simulationpara}. The ground truth SWM channel model is used for channel generation regardless of the distance, assuming that the Tx position and orientation angles remain fixed while the Rx is located arbitrarily. We consider both the Tx and Rx ULAs aligned along the $\mathrm{Z}$-axis, with propagation occurring along the $\mathrm{X}$-axis. During offline training, results for each distance are averaged over $R_\offsupsc$ orientations using a random orientation angle around the $\mathrm{Y}$-axis (represented by Euler angle $\dot{\beta}\in\mathcal{U}[-30,30]$~\cite{tarboush2021teramimo}), with $E_\offsupsc$ trials for each orientation.

Fig.~\ref{fig:prop_eta_ana} illustrates the selection metric $\eta$ as a function of distance for different signal-to-noise ratio (SNR) values with the main aim to investigate how modifying the size of Tx/Rx arrays and their structure, specifically the number of SAs and AEs, will influence the metric behavior. As expected, the proposed metric demonstrates a decreasing trend with distance. Notably, there is a considerable variation in $\eta$ in the near-field region, followed by a sharp decrease until it stabilizes at almost constant values at large distances. Fig.~\ref{fig:res1a} indicates that, with a fixed Tx configuration, enlarging the relative distance between the center SAs while keeping the Rx aperture constant, expands the near-field region. This expansion is due to the distinct regions being observed by different SAs up to a certain point. Moreover, Fig.~\ref{fig:res1b} demonstrates that increasing the Tx SA size extends the near-field region, with the proposed metric consistently capturing the relevant information, underscoring the robustness of the proposed metric. Lastly, Figs.~\ref{fig:res1c} and~\ref{fig:res1d} reveal that increasing either the Tx or Rx aperture leads to a shift in the boundary between the two fields, and the metric adeptly captures this shift.
\vspace{-4mm}
\begin{table} [htb]
\footnotesize
\centering
\caption{Simulation parameters}
\begin{tabular} {|c || c|}
 \hline
 Parameters & Values\\ [0.5ex] 
 \hline
 \hline
 Operating frequency $f_c$ & $\unit[0.3]{THz}$ \\
 System bandwidth $B_\sysidx$ & $\unit[10]{GHz}$\\
 Number of subcarriers $K$ & $16$\\
 Number of channel paths $L$& $3$\\ 
 Tx/Rx SAs $Q_\txidx/Q_\rxidx$ & $\{1, 2, 4, 8\}/\{4, 8, 16\}$ \\
 Tx AEs $\bar{Q}_\txidx$ & $\{128,256,512, 1024\}$\\
 Rx AEs $\bar{Q}_\rxidx$ & $\{8, 16, 32, 64\}$\\
 AEs/SAs spacing $\{\delta_\txidx=\delta_\rxidx\}/\{\Delta_\txidx,\Delta_\rxidx\}$&$\{\frac{\lambda_c}{2}\}/\{\bar{Q}_\txidx \delta_\txidx,\bar{Q}_\rxidx \delta_\rxidx\}$\\
  Number of training beams $M_\txidx/M_\rxidx$ & $ \frac{\bar{Q}_\txidx}{4}/\bar{Q}_\rxidx$\\
  Offline orientations/trials $\!R_\offsupsc/\!E_\offsupsc$& $31/100$\\
  Online orientations/trials $\!R_\onsupsc/\!E_\onsupsc$& $31/30$\\
 \hline
\end{tabular}
\label{table:simulationpara}
\end{table}

To examine the impact of historical observations on decisions, we assume the Rx moves across regions at a speed of $v=\unit[1]{m/sec}$, resulting in $T_{\coh}=\unit[0.42]{ms}$. We simulate a Tx and Rx of $Q_\txidx=Q_\rxidx=4$ SAs with $\bar{Q}_\txidx=256$ and $\bar{Q}_\rxidx=16$ AEs. We determined the distance, corresponding to the optimal threshold, $d_{\gamma^\opt_{\mathrm{N}\text{-}\mathrm{F}}}=\unit[10]{m}$ by simulating the normalized mean square error of suitable estimation methods (results omitted due to space limitations) and selecting the intersection of these two curves to get the best estimation accuracy. Fig.~\ref{fig:SV_vs_MV} shows the success rate of three methods - hidden Markov model $U=4$, majority vote $U=4$, and $U=1$~\cite{tarboush2023cross} - across different regions. Figs.~\ref{fig:resmv1} and~\ref{fig:resmv2} illustrate that the HMM approach increases the success rate, especially in the low SNR regime for both near- and far-fields.
\section{Conclusion}
This paper introduces a strategy to address the cross-field channel problem in an UM-MIMO sub-THz/THz-band system employing an AoSA architecture. The proposed solution involves collecting a set of decisions of the introduced selection metric and applying a HMM-based decision strategy. This strategy ensures an improved success rate in determining whether the current user is in the near- or far-fields even at low-received SNR levels.  
\newpage
\vfill\pagebreak
\balance
\bibliographystyle{ieeetr}
\bibliography{abbrev,bibliography}

\end{document}